\documentclass[epj]{svjour}
\usepackage[english]{babel}
\usepackage[latin1]{inputenc}  
\usepackage[dvips]{graphicx}
\usepackage{afterpage}

\begin{document}

\title {Freeze-out volume in multifragmentation - dynamical
simulations}

\author{
M.~P\^arlog\inst{1,2}\thanks{\emph{Correspondence to:} parlog@ganil.fr}
\and G.~T\u{a}b\u{a}caru\inst{3,2}\thanks{\emph{Present address:} Cyclotron 
Institute, Texas A\&M University, College station, Texas 77845, USA} 
\and J.P.~Wieleczko\inst{1}
\and J.D.~Frankland\inst{1}
\and B.~Borderie\inst{3}
\and A.~Chbihi\inst{1}
\and M. Colonna\inst{4,5}
\and M.F.~Rivet\inst{3}
}
\institute{
GANIL, CEA et IN2P3-CNRS, B.P.~5027, F-14076 Caen cedex, France.
\and National Institute for Physics and Nuclear Engineering,
RO-76900 Bucharest-M\u{a}gurele, Romania.
\and Institut de Physique Nucl\'eaire, IN2P3-CNRS, F-91406 Orsay
 cedex, France
\and Laboratori Nazionali del Sud, via S. Sofia 44, I-95123 Catania, Italy.
\and Physics-Astronomy Department, University of Catania, Italy. 
}
\date{\today} 
\abstract{
Stochastic mean-field simulations for multifragmenting sources at the same
excitation energy per nucleon have been performed. The freeze-out volume, a 
concept which needs to be precisely defined in this dynamical approach, was 
shown to increase as a function of three parameters: freeze-out instant, 
fragment multiplicity and system size. 
}
\PACS{
21.30.Fe;25.70.-z;25.70.Lm;25.70.Pq;24.60Ky; 
NUCLEAR REACTIONS $^{119}$Sn($^{129}$Xe, $X$), $E$ = 32 AMeV; 
$^{238}$U($^{155}$Gd, $X$), $E$ = 36 AMeV; central collisions; 
multifragmenting sources; stochastic mean-field simulations; dynamical 
evolution of the freeze-out volume.
}

\maketitle
\section{Introduction}
Several hundreds of nucleons may be brought into interaction in central heavy 
ion collisions around the Fermi energy. Such reactions are good candidates to
provoke a liquid-gas type phase transition - conceivable given the nature of 
the nuclear force - and to break the system into massive fragments 
\cite{ber83,mor02}. 
The volume of such a source of ejectiles at the 
instant when all of them become free of the attractive force and feel 
only the Coulomb repulsion - $the freeze-out$ $volume$ - brings information on 
the coexistence of phases. It is a key quantity \cite{bor02} to be connected to 
the physical observables, asymptotically measured. 
If in the statistical models \cite{gro90,bon95,lop89,rad02} it is a 
basic $a priori$ hypothesis, in the following dynamical treatment this volume 
is provided - at a given available energy - as a family of results illustrating 
the temporal and spatial evolution of the source in multifragmentation. 

Nuclear multifragmentation may occur when the source has expanded through the 
spinodal region of negative compressibility \cite{ber83} of the liquid-gas 
coexistence domain, a scenario valid for other many-body systems
too \cite{cho03,col04}. Related Stochastic Mean Field (SMF) approaches 
\cite{ayi88,ran90,cho91} consider, under different approximations, the 
amplification of density fluctuations, due to N-body correlations, by the 
unstable mean field leading to spinodal decomposition. The Brownian One-Body 
(BOB) dynamics version, simulating the fluctuations by means of a brownian force 
in the mean field \cite{cho94,gua96,gua97}, coupled to Boltzmann-Nordheim-Vlasov 
(BNV) one-body density calculations 
\cite{gua96}, was successfully confronted 
with multifragmentation data measured with the 4$\pi$ multidetector INDRA 
\cite{pou95i}. 
Two systems at close available energy per nucleon $\sim$ 7 MeV, were 
experimentally studied: 32 AMeV $^{129}$Xe + $^{nat}$Sn and 
36 AMeV $^{155}$Gd + $^{nat}$U 
\cite{sal97,mar98,fra01ii,bor01,hud03,tab03,tab00}.
The comparison between simulated events for central collisions - filtered 
according to the experimental features of INDRA 
- and the experimental data was 
recently extended from fragment multiplicity $M$, charge $Z$, largest charge 
$Z_{max}$ and bound charge $Z_{bound} = \Sigma Z$ distributions \cite{fra01ii} 
to charge \cite{bor01,tab03} and velocity correlation functions and energy 
spectra \cite{tab00}. The theoretical approaches developed in relation with 
the advanced experimental methods of nuclear physics may be esteemed in 
the new and more general physics of the phase transition in finite systems 
\cite{bor02,cho03,gul03}.

As a reasonably successful dynamical description of the multifragmentation 
process at intermediate bombarding energies, the above mentioned three 
dimensional (3D) SMF simulations provide a well adapted framework to address 
the question of the freeze-out volume. We focus in the present paper on 
disentangling the time, fragment multiplicity and source size of the 
freeze-out volume dependence. A pragmatical definition of the freeze-out 
instant is proposed.
\section{Expanding sources provided by BOB simulations}


SMF simulations of nucleus-nucleus collisions based on the Boltzmann-Langevin 
equation were proposed to treat unstable systems 
\cite{ran90,cho91,bur91} but their application to 3D nuclear collisions is  
prohibited by computational limitations. Instead, the spinodal decomposition of 
two diluted nuclear sources, mentioned above, at the same temperature 
($\approx$ 4 MeV), was mimicked through the 
BOB dynamics \cite{cho94,gua97,fra01ii}, applicable to locally equilibrated 
systems. The brownian force employed in the kinetic equations is grafted on to 
the one-body density evolution, calculated in a 
BNV approach 
\cite{gua96}, at the moment of maximum compression 
$t_0 \approx$ 40 fm/$c$ after the beginning of the collision, before the 
entrance of the system into the spinodal region. Its strength can be tuned to 
describe the growth of the most important unstable modes, ascribed to 
the density fluctuations, which need a short, but finite time to develop.   
The dispersion relation \cite{ayi95} puts them in evidence. It includes quantal 
effects and connects the characteristic time to its associated multipolarity. 
In turn, the multipolarity of the unstable collective modes, increasing with the 
size of the source, may be related to the fragment multiplicity \cite{ayi95,col02}. 
The delimitation between fragments - the "liquid drops" -  and light clusters 
- "the gas" - in which they are embedded is given by a cut-off value 
$\rho_{min} \ge 0.01 fm^{-3}$ of the nuclear density $\rho$ \cite{fra01ii}.

The ingredients of the simulations, corresponding to zero impact parameter, as well 
as the selection criteria for the complete events from experimental central  
collisions, can be found in \cite{fra01ii,fra01i}. 
The fragments are 
defined as having the atomic number $Z \ge$ 5. In the reported calculation, as in 
the experimental selection, only events having the final fragment multiplicity 
$M \ge$ 3 were considered \cite{fra01ii}. 
The calculated total charge 
of the multifragmenting sources in the spinodal zone, $Z_{tot}$ = 100 
for $^{129}$Xe + $^{119}$Sn and $Z_{tot}$ = 142 
for $^{155}$Gd + $^{238}$U 
are close to the experimentally 
reconstructed ones \cite{fra01ii}; 
INDRA identifies the mass 
of the light charged products ($Z < 5$) but not that of the fragments.  
The Skyrme force used in our simulations is not isospin 
dependent. Consequently, the mass numbers of the sources: $A_{tot}$ = 238 
(for the light one) and $A_{tot}$ = 360 (for the heavy one) 
correspond to the conservation of the entrance channel $N/Z$ ratio. The charge 
distributions (normalized to the fragment multiplicity of the event) measured 
\cite{riv98} or simulated in 
BOB calculations \cite{fra01ii}, are identical in the 
two cases  - consistent with a bulk effect in the involved multifragmentation 
process. 

\begin{figure}[!hbt]
\includegraphics*[width=8.5cm]{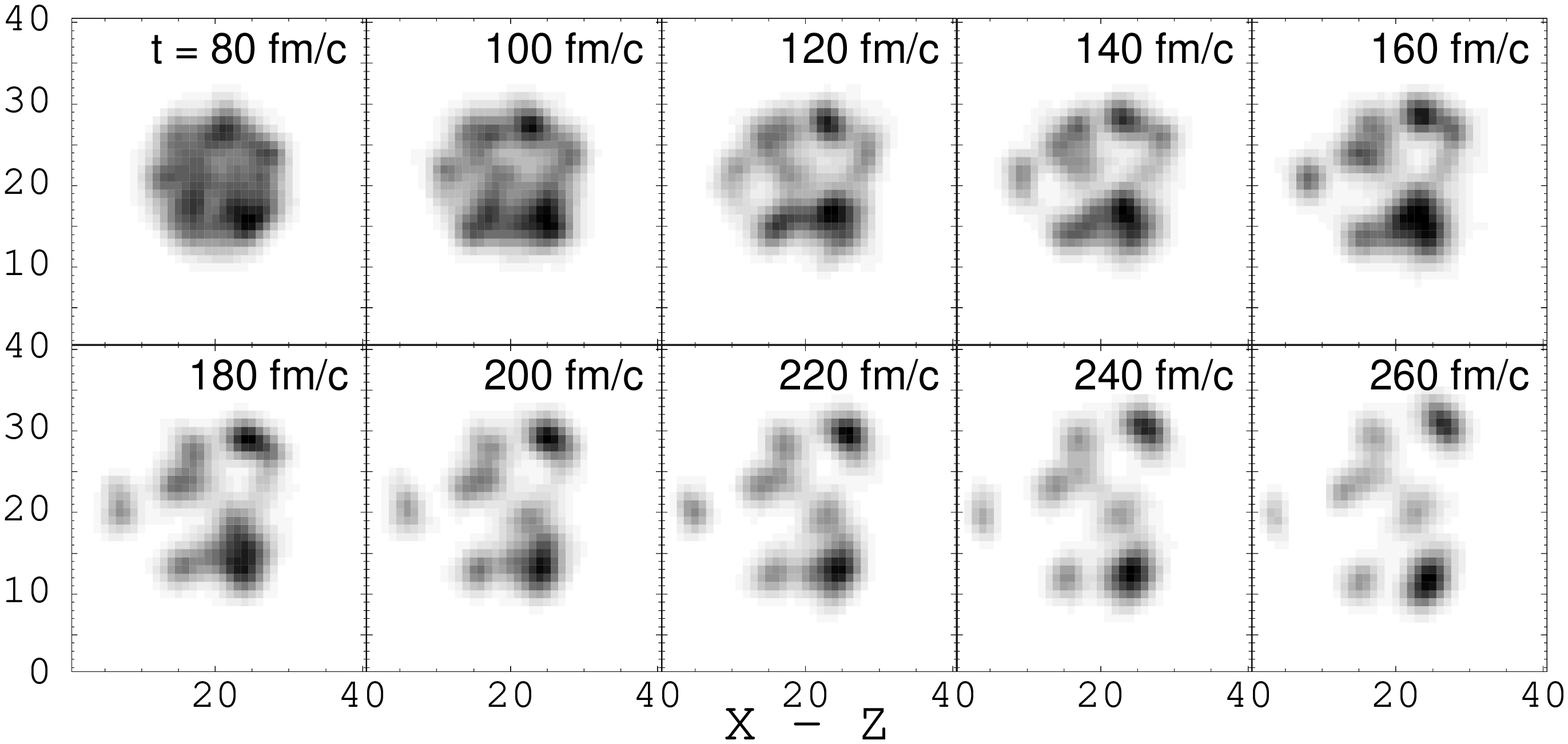}
\includegraphics*[width=8.5cm]{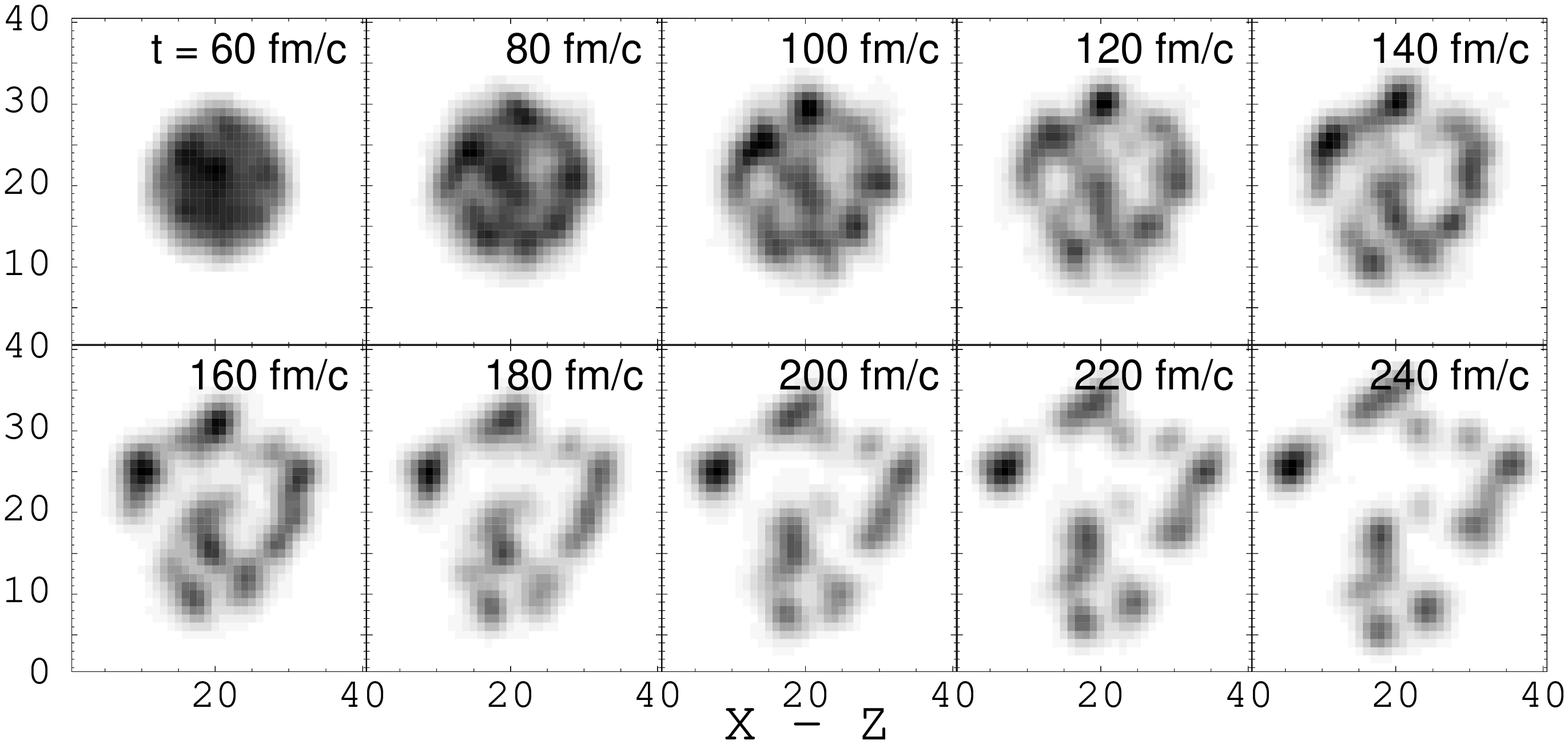}
\caption{\footnotesize One event density evolution for each of the two 
multifragmenting sources: $A_{tot}$ = 238 - upper panel - and $A_{tot}$ = 360 - 
lower panel. The collision direction in the entrance channel is along Z axis 
and the unit is fm on both axes of these X - Z views. The centre of mass 
coordinates are X = Z = 20 fm.}
\label{fig10}
\end{figure}

Starting from the two initial partners of the reaction, the BNV calculation 
leads to a spherical distribution of matter. It undergoes a self-similar 
expansion, which is not 
dramatically altered by the BOB simulations - 
Fig. \ref{fig10}. 
The simulated sources continue their rather isotropical 3D expansion in time, 
no particularly elongated or flat shapes being produced. The concept of a 
radially symmetric freeze-out volume keeps its full sense, provided that the 
related instant may be singularized.  
\subsection{The freeze-out instant definition}
 
\begin{figure}[!hbt]
\includegraphics*[width=9.cm]{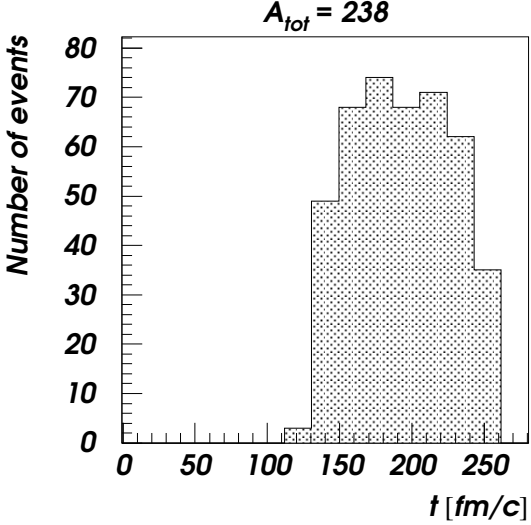}
\caption{\footnotesize The distribution of the moments at which the final 
multiplicity $M$ = 5 was reached for the source having $A_{tot}$ = 238.}
\label{fig1}
\end{figure}

The fragments are not all formed at the same moment, their mean 
multiplicity increasing in time, up to $\sim$ 250 fm/$c$ 
\cite{fra01ii} 
when it saturates at a value of about 5 for 
the lighter source ($Z_{tot}$ = 100, $A_{tot}$ = 238) associated to the 
Xe + Sn reaction, and of about 8 for the heavier one 
($Z_{tot}$ = 142, $A_{tot}$ = 360) associated to the Gd + U reaction. 
Indeed, even for the same final multiplicity of a source, there are events 
where the density fluctuations grow up faster and others where the process is 
slower. A question of definition appears related to the freeze-out instant. 
Our BOB calculations are recorded in steps of time $\Delta t$ = 20 fm/$c$ 
starting from $t_0$. Each event is traced back in steps of 20 fm/$c$ from its 
asymptotical configuration ($t \approx$ 250 fm/$c$) of final multiplicity 
$M$ to the moment when the fragment multiplicity decreases one unit.
We define the freeze-out instant of an event as the moment when its $final$ 
fragment  multiplicity $M$ was established. It means that if at 
$t_{i - 1} = t_0 + (i-1) \Delta t$ the event has the multiplicity $M - 1$ and 
at $t_i = t_0 + i \Delta t$ it reaches the final multiplicity $M$, $t_i$ will 
be considered as freeze-out instant. All events which got their final 
multiplicity $M$ at the moment $t_i$ are treated together. Fig. \ref{fig1} 
shows, as an example, the distribution of the moments $t = t_i$ at which the 
final multiplicity $M$ = 5 was reached for the lighter source $A_{tot} = 238$: 
$t \in [120, 260]$ fm/$c$.
\subsection{Freeze-out configurations}

\begin{figure}[!hbt]
\includegraphics*[width=8.cm]{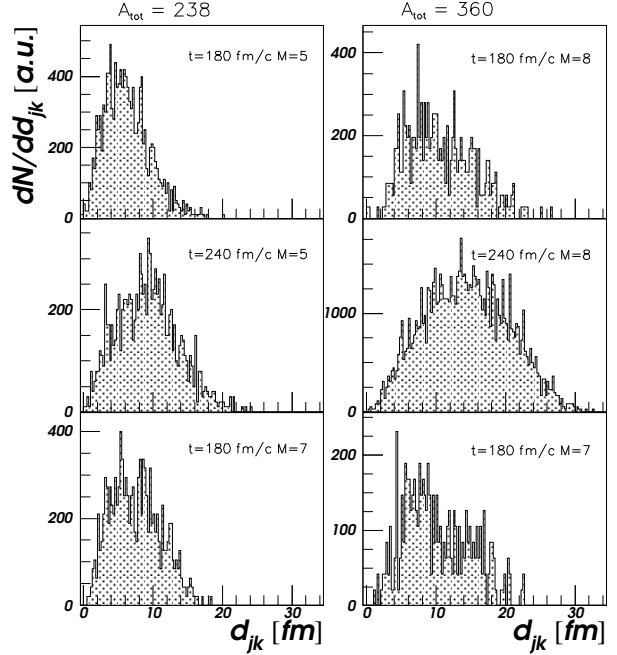}
\caption{\footnotesize The distributions of relative distance $d_{jk}$ between 
the surfaces of two fragments $j,k$, simulated event by event and  
related to two sources: $A_{tot}$ = 238 - left column - and 
$A_{tot}$ = 360 - right column, for various freeze-out instants and
multiplicities.}
\label{fig2}
\end{figure}

For the sake of simplicity, the fragments are considered 
as spheres of normal density, but the results are quite independent of 
this particular hypothesis. The relative distance $d_{jk}$ between the surfaces 
of two fragments $j,k$ ($j,k = 1 \div M$ and $j \neq k$) in one event, 
that will be called the intra-fragment distance in the following, is much 
varying for a given final multiplicity $M$: from a minimum 
of the order of 1 fm between the two most recently splitted fragments 
to a maximum value which increases with $t$.  
This behaviour is shown - for the source (100,238) and the
multiplicity $M$ = 5 
- in the left-upper and left-middle panels of Fig. \ref{fig2}, at two 
different freeze-out instants: 180 and 240 fm/$c$. 
The distribution evolves towards larger distance values in time, from an 
asymmetric to a more symmetric shape.
For the same two instants, the right-upper and right-middle panels of  
Fig. \ref{fig2} present the distributions concerning the source (142,360) and 
$M$ = 8. 
The evolution is similar to the lighter source case. Together with the lower 
panels, related to the freeze-out instant $t$ = 180 fm/c and the multiplicity 
$M$ = 7 for both sources, the upper graphs let also one see that, at the same 
moment ($t$ = 180 fm/c), the 
mean value of the relative distance depends on $M$ for a given source,
while for different sources and the same multiplicity, e.g. $M$ = 7, 
on the size of the source. 

The distribution of the intra-fragment distances at a 
certain moment is quite illustrative for the momentary spatial configuration 
of the source: the shortest distances concern the first order neighbours, the 
longest ones the largest size of the sources and, above all, the intermediate 
values are qualitatively informing about the dilution of the source. The first 
and second order moments of the distribution and their dynamical evolution are 
appropriate to synthetize this last aspect. As expected, these quantities, 
analysed multiplicity by multiplicity, are increasing with increasing $t = t_i$, 
testifying on the enlargement of the matter distribution in the nuclear source 
in time. These values are higher for the heavier source than for the lighter 
one, and their increase in time seems to be slightly more 
pronounced. The fact has to be put in connection with the total Coulomb 
repulsion and the radial flow, higher for the heavier system than for the 
lighter one \cite{fra01ii,tab00}. An interesting result is that, at constant 
values of $t = t_i$, the mean and the variance of intra-fragment distance 
distributions increase with the multiplicity - a little bit more accentuated 
for the lighter system. The slope of this dependence is higher at later moments
$t$.    

\begin{figure}[!hbt]
\includegraphics*[width=8.cm]{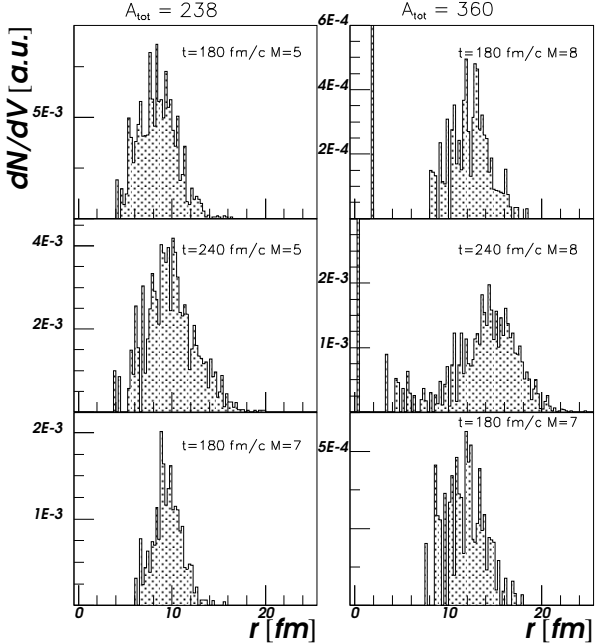}
\caption{\footnotesize  Radial distributions of the fragment 
concentration, simulated event by event and related to two sources: 
$A_{tot}$ = 238 - left column - and $A_{tot}$ = 360 - right column, for various 
freeze-out instants and multiplicities.}
\label{fig3}
\end{figure}

For each group of events reaching a final multiplicity $M$ at a certain moment 
$t = t_i$, one may look for the corresponding local fragment concentration: 
d$N$/d$V$ as a function of the vector radius absolute value $r$ of the 
fragment position in the source reference framework. 
Examples of such distributions (at $t$ = 180 and 240 fm/$c$) are given in the 
left-upper and left-middle panels of Fig. \ref{fig3} for the lighter source   
and $M$ = 5. 
When the same source and final multiplicity are involved, 
the mean radius $\bar{r}$ and full width at half maximum ($FWHM$) increase with 
the time $t = t_i$ when the last two fragments are separated. In the 
right-upper and right-middle panels of Fig. \ref{fig3} are represented - for 
the heavier source  
- the local fragment 
concentration distributions at these two moments and the final multiplicity 
$M$ = 8. As previously, the distribution evolves in time towards larger 
radii; these radii are longer than in case of the lighter source.  
For the same final multiplicity: $M$ = 7 and the same moment: $t$ = 180 fm/$c$ 
- the distributions represented in the lower panels for both systems - the 
mean radius $\bar{r}$ and $FWHM$ are longer in case of the heavier source. 
The upper and the lower pannels together show that, for a given source at 
the same freeze-out instant, $\bar{r}$ varies in the same sense as $M$.

Except the lowest multiplicity case for the lighter source, the radial 
distributions of the fragment concentration at the freeze-out instant are 
practically empty towards short radii, as shown in  Fig. \ref{fig3}. They may 
be interpreted hence as reminiscences of bubble-like configurations.
\section{The freeze-out volume}

\begin{figure}[!hbt]
\includegraphics*[width=8.cm]{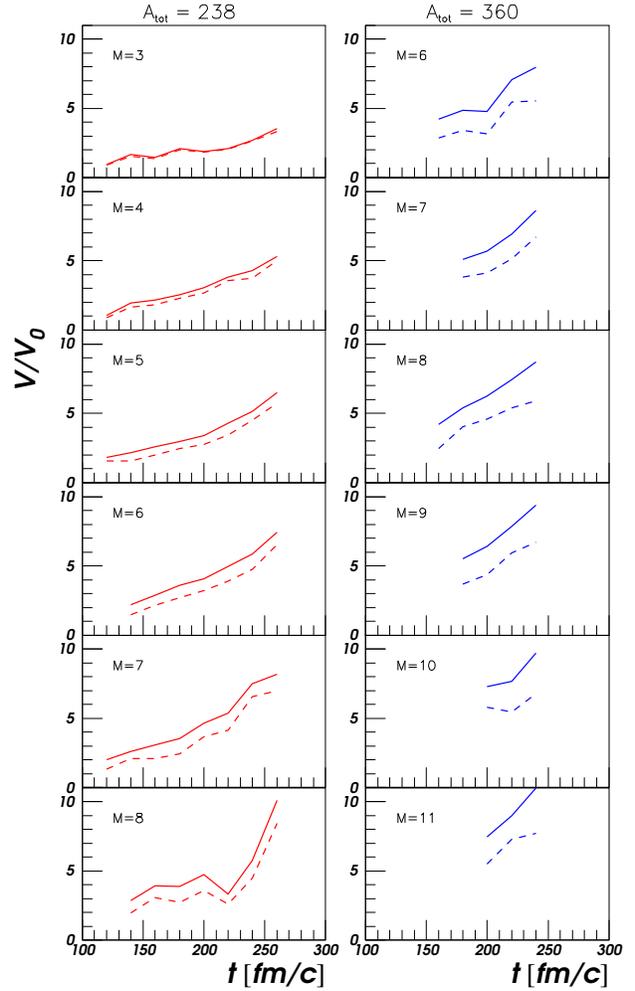}
\caption{\footnotesize  Freeze-out volumes as a function of the freeze-out 
instant, for two sources: $A_{tot}$ = 238
- left column - and $A_{tot}$ = 360 
- right column. 
The full curves concern full spheres, while the dashed ones correspond to 
hollow spherical envelopes (see the text for explanations).}
\label{fig6}
\end{figure}

Once the freeze-out instant defined and the configurations when the latest 
formed fragments get free of nuclear interaction found, one may look for the
corresponding volume. A sphere of radius $\bar{r} + FWHM/2$ englobes most of 
the fragment centres. Its volume - a good estimate of the quantity of interest 
- normalized at the volume $V_0 = (1.2)^3 A_{tot}$ fm$^{3}$ of the source at 
normal density $\rho_0$, is considered as a function of the freeze-out instant 
$t = t_i$ at constant final multiplicity $M$ of fragments. It is represented 
with full curves in the left column panels of Fig. \ref{fig6} for the  
source $A_{tot}$ = 238. 
The expansion of the source in time is evidenced: the dynamic process 
delivering the same final number $M$ of fragments implies higher volumes if 
it takes place at later instants. These volumes increase from top to bottom 
with $M$, as well as the slopes of their variation. By adding to the above 
spheres the complements of fragment volumes which are exceeding them, one 
gains roughly 10$\%$ on the freeze-out values, but the general behaviour is 
not changed.     

In fact, as shown in Fig. \ref{fig3}, the shapes of the fragment concentration 
distributions at freeze-out are generally gaussian like. About 
75$\%$ of their integral is hence comprised between $\bar{r} - FWHM/2$ and 
$\bar{r} + FWHM/2$. The hollow envelopes delimited by the spheres with these 
radii are a kind of lower limits of the corresponding freeze-out volumes;
always normalized at $V_0$, they are drawn as dashed curves. The right column 
panels of Fig. \ref{fig6} show similar results for the source $A_{tot}$ = 360. 
The curves seem to rise slightly more rapidly than in the left column, in 
relation with the Coulomb effect and the radial flow.  

\begin{figure}[!hbt]
\includegraphics*[width=8.cm]{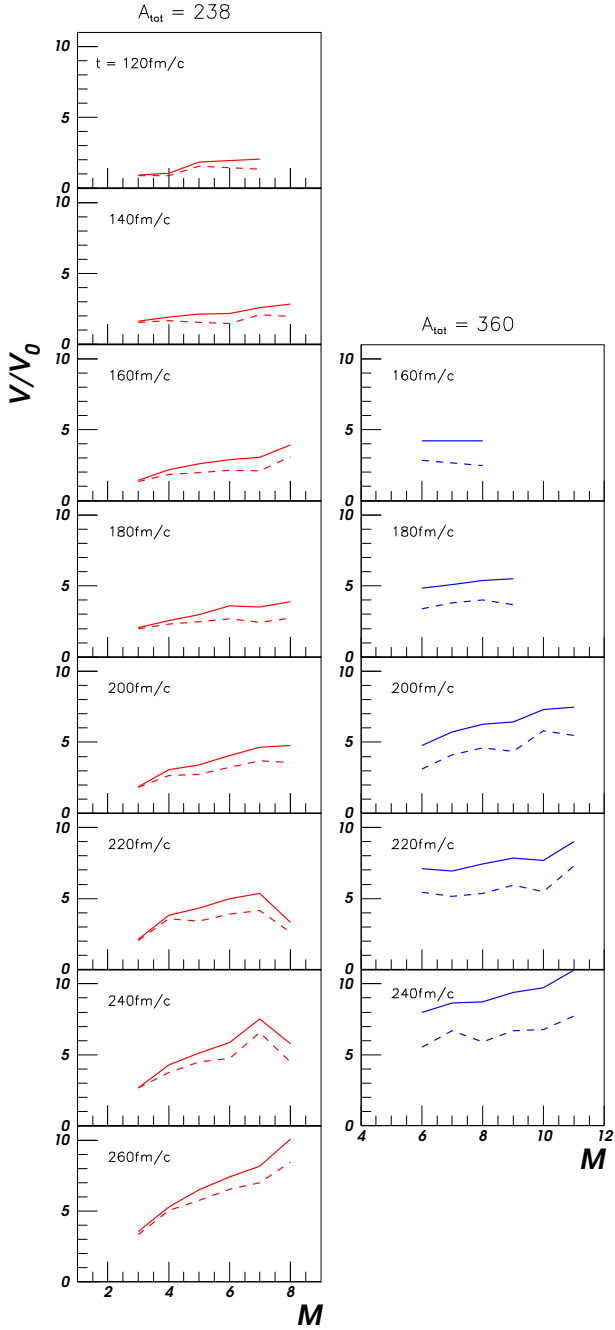}
\caption{\footnotesize  Freeze-out volumes as a function of multiplicity, at
different freeze-out instants, for two sources $A_{tot}$ = 238
- left column - and $A_{tot}$ = 360
- right column. 
The full curves concern full spheres, while the dashed ones correspond to 
hollow spherical envelopes (see the text for explanations).}
\label{fig7}
\end{figure}

The evolution with the multiplicity of the freeze-out volume defined above, 
calculated for full or hollow spheres and normalized at the volume of the 
corresponding source at normal density, is shown in Fig. \ref{fig7} at given
freeze-out moments. 
The dilution of the source increases with the fragment multiplicity - slightly 
faster for the lighter system, presenting a higher relative variation of $M$ 
(left column panels) than for the heavier one (right column panels); the 
variation is more pronounced at larger times. It reflects the 
mechanism considered here for the multifragmentation: the density fluctuations.
The separation of the fragments in such an expanding source is reached on 
behalf of the lower density domains: the larger the number of fragments, the
larger the number of zones between them. Consequently, at the same freeze-out 
instant, the higher the multiplicity, the bigger the source volume. From the 
energetic point of view, a higher final fragment multiplicity $M$ implies a 
higher fraction of the excitation energy consumed as binding (mainly surface) 
energy. 

The freeze-out volumes provided by the present 
BOB calculation, averaged
over time and multiplicity, are, of course, well fitting with those extracted, 
in the same framework, by using the mean multiplicity information. 
The corresponding densities $\rho$ are compatible with the general prediction 
$\rho_0/10 < \rho < \rho_0/2$ of the Statistical Model for Multifragmentation
(SMM) \cite{bon95}, in particular with the value $\rho_0/3$ used to study - 
in a nonsphericity hypothesis in this latter model - the fragment velocity 
correlations in the 32 AMeV $^{129}$Xe + $^{nat}$Sn system \cite{sal97}. The 
present density values are lower than the average densities corresponding to 
the same domain of excitation energy, extracted from nuclear caloric curves 
\cite{nat02,sob04}. The microcanonical model, analysing - all multiplicities 
together - the same multifragmenting systems \cite{rad02}, leads to volume 
values close to the present results but with a slightly lower dilution for the 
heavier system than for the lighter one.

The increase of the freeze-out volume with the basic experimental 
observable which is the fragment multiplicity in an event is 
more pronounced than that obtained when compacity criteria are used to fill 
the freeze-out domain in a static scenario. 
For the first time, reliable values 
as a function of time and final multiplicity are calculated. They give a 
dynamical image of the multifragmentation process at Fermi incident energies. 

The final multiplicity of fragments is a measurable observable, while the time 
information is not directly accessible in the experiments. By weighting the 
freeze-out volume values got at different freeze-out instants with respect to 
the corresponding number of events, one may obtain a kind of average freeze-out 
volume at a given final multiplicity. This "time" averaged quantity, which 
\begin{figure}[!hbt]
\includegraphics*[width=8.cm]{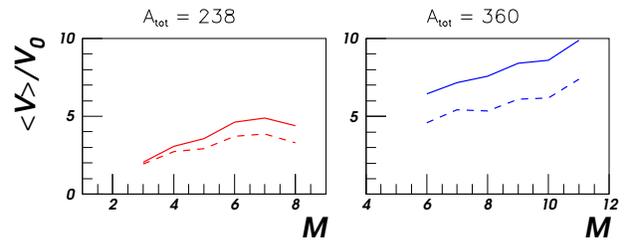}
\vspace{-14.6cm}
\caption{\footnotesize As in Fig. \ref{fig7} but for "time" averaged freeze-out 
volumes; see the text for explenations. 
}
\label{fig8}
\end{figure}
looses much of its physical signification, indeed, is rapidly rising with $M$, 
as shown for the two sources in Fig. \ref{fig8}. 
The variety of fragment configurations, connected to the dynamical 
fluctuations of the system volume on the fragmentation path, has been recently 
related to monopole oscillations \cite{col04}. They may push the system towards 
a metastable configuration which eventually recontracts leading to low fragment 
multiplicities, or develops into a hollow configuration which fragments at 
higher multiplicities.   
\section{Conclusions}

Stochatisc mean field simulations concerning the multifragmenting 
sources formed in 32 AMeV $^{129}$Xe + $^{119}$Sn and 36 AMeV 
$^{155}$Gd + $^{238}$U reactions have been performed. Their validity has already
been checked 
\cite{fra01ii,bor01,tab03,tab00} by reproducing measurable physical 
observables determined for these reactions with the 4$\pi$ multidetector INDRA. 
We found that the moment of separation of the latest two nascent fragments - 
the definition adopted for the freeze-out instant - is mainly distributed in the 
range $\sim (150 - 250)$ fm/$c$; multifragmentation is a dynamical process which 
is fast but needs a finite time. The topology of the associated freeze-out 
configurations are more complex but also more realistic as compared to the 
simplifying hypothesis in which the fragments, separated by a distance of the 
order of the nuclear interaction range, are forced to fill in a prescribed 
volume, as generally done in statistical codes. 

The corresponding freeze-out volume could thus be disentangled 
in connection with the freeze-out instant, final fragment multiplicity and 
source size. For a given source and a given multiplicity, the evolution of 
this quantity illustrates the continuous expansion of the source in time. 
On the other hand, at the same freeze-out instant, the volume of one source 
increases with the fragment multiplicity, a basic measurable observable. As 
part of the energy associated to the internal degrees of freedom of the source 
is going along into fragment separation energy, its volume is dramatically 
increasing. And finally, for the same moment and the same multiplicity, the 
freeze-out volume is bigger for the heavier source, involving higher Coulomb 
repulsion and radial flow. 

The dilution of a multifragmenting source, quantified as the ratio between 
its volume at freeze-out and its volume at normal density, is therefore 
increasing with time, fragment multiplicity and source size. Further SMF 
simulations, employing an isospin dependent nuclear force, would allow to 
study the isospin fractionation \cite{bar98,cho99}, a phenomenon expected to 
be amplified at high source dilution. A comparative experimental investigation 
of isoscaling characteristics of these two systems prepared at the same 
available energy is desirable too.

\end{document}